\documentclass[preprintnumbers,twocolumn,aps,amsmath,amssymb,showpacs,nofootinbib,superscriptaddress]{revtex4-2}

\usepackage{amsmath}
\usepackage{graphicx,enumerate,xcolor,enumitem,xspace,braket,mathrsfs} 
\usepackage[colorlinks,citecolor=blue]{hyperref}
\usepackage{comment}
\usepackage{float}
\usepackage{booktabs}
\usepackage{cleveref} 
\usepackage{braket}

\usepackage[normalem]{ulem}

\usepackage{tikz}
\usetikzlibrary{intersections}
\usetikzlibrary{positioning}
\usetikzlibrary{decorations.pathmorphing}
\usetikzlibrary{decorations.markings}
\tikzset{snake it/.style={decorate, decoration=snake}}
\usepackage{tikz-feynman}

\definecolor{nicegreen}{rgb}{0.,0.5,0.}

\begin{document}
\preprint{PITT-PACC-2620}

\title{Chains of Vector-like Leptons at Colliders}

\author{Tao Han}
\email{than@pitt.edu}
\affiliation{PITT PACC, Department of Physics and Astronomy,\\ University of Pittsburgh, 3941 O’Hara St., Pittsburgh, PA 15260, USA}

\author{Tong Arthur Wu}
\email{tow39@pitt.edu}
\affiliation{PITT PACC, Department of Physics and Astronomy,\\ University of Pittsburgh, 3941 O’Hara St., Pittsburgh, PA 15260, USA}

\begin{abstract}
Vector-like leptons (VLLs) are predicted in the chain flavor models, which possess characteristic features to explain the mass hierarchy and mixing of the fermions in the Standard Model (SM). 
We investigate the production and decay of the VLL chains and quantify the discovery potential at the LHC, FCC-hh, and lepton colliders. The resonant mass sets the new physics scale responsible for the fermion mass generation. The observation of the VLL multiplets associated with each lepton flavor yields the chain structure. The mass difference among the VLL multiplets determines the hopping parameter that gives the power-law pattern for the observed SM fermion masses. The single VLL production via vector boson-lepton fusion directly measures the partial width, equivalently the new Yukawa coupling and the VLL mass hopping parameter. At hadron colliders, the leading production mechanism is pair production via their gauge interaction. We find that HL-LHC may have the sensitivity to reach the chain scale of $800-900\,\mathrm{GeV}$ and FCC-hh reaches $>2.5\,\mathrm{TeV}$. In $e^+e^-$ and $\mu^+\mu^-$ colliders, VLL pair production is a clear discovery channel once the threshold is reached.
 Single production via $Z$-lepton fusion becomes more accessible owing to a lower mass threshold and extends signal discovery to the collider kinematic limit. 
\end{abstract}

\maketitle
\begingroup
\hypersetup{linkcolor=black}
\tableofcontents
\endgroup

\section{Introduction}

Understanding the mechanism for fermion mass generation and their mixing pattern remains a great challenge in and beyond the Standard Model (SM) in particle physics. Although Yukawa interaction can provide an adequate fit to the observation, it leaves the hierarchical masses and mixings among the three generations unexplained. Novel ideas have been proposed in the construction of viable flavor models, such as the Froggatt-Nielsen mechanism \cite{Froggatt:1978nt}, in which approximate flavor symmetries generate hierarchical Yukawa couplings through powers of a symmetry-breaking parameter for different generations. 
Related approaches organize interactions along chains, as in dimensional deconstruction \cite{Arkani-Hamed:1999ylh,Grossman:1999ra,Gherghetta:2000qt,Kaplan:2001ga,Arkani-Hamed:1999pwe,Bordone:2017bld,Fuentes-Martin:2022xnb} 
and the clockwork mechanism \cite{Giudice:2016yja,Patel:2017pct,Ibarra:2017tju,Alonso_2018,AbreudeSouza:2019ixc,Hong:2019bki,vonGersdorff:2020ods,Kang:2020cxo,Babu:2020tnf,Altmannshofer:2021qwx,Mohanta:2026jnq,Ibarra:2025eya}. 
However, the existing stringent limits from precision flavor physics experiments push the new physics scale rather high, typically exceeding  100 TeV, rendering the new physics effects not directly observable at colliders \cite{Isidori:2010kg,Calibbi:2012at,Ishiwata:2015cga,Silvestrini:2018dos,Smolkovic:2019jow,Aebischer:2020dsw}. 

Much effort has been put into considering flavor physics models that are not far above the electroweak scale \cite{Leurer:1992wg,Buras_2011,Greljo:2025mwj}. Recently, a model has been proposed in which fermion mass hierarchies arise from chains of SU(2)$_L$-singlet vector-like quarks and charged leptons linked by soft flavor-symmetry-breaking mass terms \cite{Arkani-Hamed:2026wwy}. Its restricted interaction structure suppresses flavor-changing and CP-violating effects, allowing the new vector-like (VL) fermions to lie near the TeV scale.

This connection between flavor hierarchies and an accessible heavy spectrum motivates direct collider tests. Vector-like quarks (VLQs) and leptons (VLLs) have been a longstanding focus of collider studies at LEP and LHC \cite{L3:2001xsz,Morais:2021ead,Kawamura:2023zuo,Ashanujjaman:2022cso,Kumar:2015tna,Koren:2021ypn,ATLAS:2025cdi,Guedes:2021oqx,Cui:2026wwo,Cao:2023smj,Baspehlivan:2022qet,Chala:2020odv,CMS:2019hsm,ATLAS:2024mrr,CMS:2024bni,ATLAS:2023sbu,CMS:2022cpe,ATLAS:2025wgc,CMS:2022nty}. In the chain model, the leading mode for VL fermion production is through gauge-mediated pair production, which is model-independent and directly determines the new physics scale responsible for the SM fermion mass generation. The heavy–light Yukawa couplings induce single VL fermion production. The production rate is directly proportional to the partial width governed by the characteristic mixing parameters and thus provides a direct probe of the underlying interaction. The VL fermions subsequently decay into SM fermions accompanied by a gauge boson $W^\pm$, $Z$, or a Higgs boson. The high-energy decay products and the resonant invariant masses of the reconstructed decay systems provide distinct kinematics from the SM backgrounds. A characteristic feature of the chain model is the presence of multiple heavy fermionic states associated with each SM fermion with appreciable mass splittings, producing rich resonant structures and amplifying the production rate. The mass difference among the VL fermionic  multiplets determines the hopping parameter that governs the power-law pattern of the fermion mass. Those patterns offer a distinctive probe of the underlying chain structure in searches for VL fermions.

In this work, we study the collider phenomenology and the discovery potential of vector-like leptons in the chain model at the HL-LHC, FCC-hh, and $e^+e^-$ and $\mu^+\mu^-$ colliders. In a hadron collider, model-independent pair production via electroweak gauge interactions dominates, while the single VLL production through flavor mixing is subdominant and model-dependent. However, it offers direct measurement of the decay width and thus the model coupling parameters. In lepton colliders, the two channels can have comparable rates, providing complementary probes of the chain structure. A notable feature is that single production is dominated by interactions with longitudinal gauge bosons, which gives rise to vector-boson-fusion (VBF) channels in hadron colliders and channels from $Z$ boson radiation in lepton colliders. We find that pair-production searches may probe the chain scale up to 0.9 (2.5) TeV at the HL-LHC (FCC-hh). At lepton colliders the reach goes as high as the collider kinematic limit for both pair and single production.

\section{A Low-scale Flavor Model}

\subsection{Chain Model Setting}

Vector-like leptons are extensions of the SM lepton sector in which the left- and right-handed components transform identically under the SM gauge group. Unlike chiral SM leptons, VLLs can acquire gauge-invariant Dirac masses independently of electroweak symmetry breaking. In the chain model, sequential VLLs are introduced to induce a series of mass mixings, providing a mechanism for generating a suppressed and hierarchical effective Yukawa coupling. For each lepton flavor (generation) $\ell$, there is a chain of VLLs, denoted by $E_i\ (i=1,2,...,D$ with multiplicity $D$). The VLLs introduced in the chain model \cite{Arkani-Hamed:2026wwy} are gauge-singlets under $\mathrm{SU}(2)_L$ and are only charged under $\mathrm{U}(1)_Y$ given by 
\begin{equation}
    E_i^c : (1,1),\hspace{1em} E_i : (1,-1) .
\end{equation} 
The VLL Lagrangian is of a chain structure and is given by 
\begin{align}
    \mathcal{L}_{\mathrm{VLL}} &= y \ell H E_1^c + M_i E_i E_i^c + m_{ij} E_i E_j^c + m E_D \ell^c + \mathrm{h.c.} \label{lag1}
\end{align}
The new Yukawa coupling $(y)$ connects the left-handed SM lepton field $(\ell)$ to the first VLL $(E_1)$. In addition to a vector-like mass $(M)$, the hopping terms of $(m_{ij}, m)$ connect the right-handed SM lepton through the chain of VLLs. In order to suppress the flavor-changing neutral currents, the chains for the three SM leptons are disconnected. In accordance with the SM lepton generations, the multiplicity $D=1,3,5$ for $\tau,\mu$ and $e$, respectively, thus achieving the hierarchical mass and mixing pattern through the chain structure \cite{Arkani-Hamed:2026wwy}.

In this work, we use the muon flavor as the primary example for illustration. The discussion and results are equally applicable for other flavors with simple rescaling and different multiplicity $D$. For simplicity, we take the equal mass terms $M_i=M$, and the equal hopping terms 
\begin{equation}
    m_{ij}=m=\epsilon M.
\end{equation}
Those hopping terms connect the nearest VLL neighbors, and the hopping mass parameter $\epsilon$ controls the fermion mass hierarchy. This minimal choice captures the essential features of the model. For the muon flavor, there are three VLL states, and the consequent Lagrangian after electroweak symmetry breaking reads \cite{Arkani-Hamed:2026wwy}
\begin{equation}
    \mathcal{L}_{\mathrm{VLL}} = \frac{1}{\sqrt{2}} y \mu h E_1^c +
    M \begin{bmatrix}
        \mu \\ E_1 \\ E_2 \\ E_3
    \end{bmatrix}^T 
    \begin{bmatrix}
        0 & \lambda & 0 & 0\\ 0 & 1 & \epsilon & 0 \\ 0 & 0 & 1 & \epsilon \\ \epsilon & 0 & 0 & 1
    \end{bmatrix}
    \begin{bmatrix}
        \mu^c \\ E_1^c \\ E_2^c \\ E_3^c
    \end{bmatrix}  + \mathrm{h.c.}  
    \label{lag2}
\end{equation} 
with the mixing parameter
\begin{equation}
    \lambda \equiv \frac{yv}{\sqrt{2}M} . 
\end{equation}
The main features of the chain model are made clear by Eq.~(\ref{lag2}). First, the Yukawa coupling $y$ induces portal coupling between a single VLL and SM particles, which predominantly contributes to single production of VLL. Second, the parameter $\lambda$ determines the mixing between the left-handed VLL and the SM lepton. Third, the hopping mass parameter $\epsilon$ shapes the splitting of the masses between different VLL states. Lastly, the masses of the SM leptons are governed by $y\epsilon^D$, leading to a hierarchical pattern. 

The chain model can also be applied to explain the smallness of the neutrino mass, with a similar Yukawa coupling of $y_N \ell H^\dagger N^c$, where $N$ is the neutral heavy lepton (HNL), and a mixed mass matrix with 5 HNL states \cite{Arkani-Hamed:2026wwy}. Due to the similar Yukawa and mixing structure, the features and production rates discussed in this work are equally applicable to the searches of HNLs in the chain model, with minimal rescaling and substitutions.

\subsection{Mass and Mixing}
\label{sec:mass and mixing}

The mass matrix in Eq.~(\ref{lag2}) generates mixings between $\mu$ and $E_i$, and among $E_{1,2,3}$ themselves. We first disentangle the light and heavy leptons through a block diagonalization, after which the light lepton goes to mass eigenstate, and we denote the heavy leptons in the new basis as flavor eigenstates $E_{i,f}$ and 
$E_{i,f}^c$.\footnote{It should be noted that those flavor or mass eigenstates are all defined with respect to a certain generation of the SM leptons. The VLLs associated with different lepton generations are unmixed.} The left- and right-handed fields undergo different rotations 
\begin{align}
    [\mu\ E_1 \ E_2 \ E_3]^T &= U_L [\mu_m\ E_{1,f} \ E_{2,f} \ E_{3,f}]^T ,\\
    [\mu^c\ E_1^c \ E_2^c \ E_3^c]^T &= U_R [\mu^c_m\ E^c_{1,f} \ E^c_{2,f} \ E^c_{3,f}]^T ,
\end{align}
where
\begin{equation}
    U_L = \begin{bmatrix}
        1 & \lambda & -\lambda\epsilon & \lambda\epsilon^2 \\
        -\lambda & 1 & 0 & 0 \\ 
        \lambda \epsilon & 0 & 1 & 0 \\
        -\lambda\epsilon^2 & 0 & 0 & 1
    \end{bmatrix},\  U_R = \begin{bmatrix}
        1 & \epsilon^3 & -\epsilon^2 & \epsilon \\
        -\epsilon^3 & 1 & 0 & 0 \\ 
        \epsilon^2 & 0 & 1 & 0 \\
        -\epsilon & 0 & 0 & 1
    \end{bmatrix} 
    \nonumber 
\end{equation}
for $\epsilon,\lambda \ll 1$. The resulting mass term reads
\begin{equation}
    \mathcal{L}_{\mathrm{VLL}} \supset \begin{bmatrix}
        \mu_m \\ E_{1,f} \\ E_{2,f} \\ E_{3,f}
    \end{bmatrix}^T 
    \begin{bmatrix}
        -\frac{1}{\sqrt{2}}y \epsilon^3 v & 0 & 0 & 0\\ 0 & M & \epsilon M & 0 \\ 0 & 0 & M & \epsilon M \\ 0 & 0 & 0 & M
    \end{bmatrix}
    \begin{bmatrix}
        \mu_m^c \\ E_{1,f}^c \\ E_{2,f}^c \\ E_{3,f}^c
    \end{bmatrix}. \label{eq:Mmatrix2}
\end{equation}
The muon therefore receives a power-suppressed mass 
\begin{equation}
    m_{\mu} = \frac{1}{\sqrt{2}}\,y\,\epsilon^3 v.
\end{equation} 
For $y\sim\epsilon\sim0.1$, one naturally generates a small effective muon Yukawa $y_\mu=y \epsilon^3 \sim10^{-4}$. 
In comparison, the electron mass will be suppressed by higher powers due to the longer chain of a higher VLL multiplicity, given by $m_e=y\epsilon^5v/\sqrt 2$. 

The flavor basis for heavy leptons is particularly interesting and useful because of the hierarchical structure of the mixing between $\mu$ and $E_{1,2,3}$, which is equivalently predicted by the Feynman diagrams of the mixing chain \cite{Arkani-Hamed:2026wwy}. With the leading mixing $\mu \sim \mu_m +\lambda E_{1,f}$,
one can immediately identify the dominant production channel.
In addition to the $U(1)_Y$ gauge interaction of the VLL, the mixing generates additional interactions for heavy VLL production. Note that there is no off-diagonal coupling $\mu E \gamma$ or $\mu_R E_R Z$. The relevant and leading interactions are given by
\begin{equation}
    \mathcal{L}_{\mathrm{int}} \supset -\lambda \frac{g_Z}{2} Z_\rho E_{1,f}^\dagger \overline{\sigma}^\rho \mu  + \lambda \frac{g_2}{\sqrt{2}} W^-_\rho E_{1,f}^\dagger \overline{\sigma}^\rho \nu_\mu + \mathrm{h.c.} \label{eq:mixedgauge}
\end{equation}
The mixing interactions involving $E_{2,f},E_{3,f}$ are simply scaled by a factor of $-\epsilon$ and $\epsilon^2$, respectively. The small mixing in $\lambda$ leads to rather weak coupling between VLLs and the SM leptons and gauge bosons, $\lambda g\sim y m_V / M$, suppressed by the ratio of the vector boson mass to the VLL mass. However, for the longitudinal gauge boson, its polarization vector at high energy $\varepsilon^\mu\sim p^\mu/m_V$ enhances the coupling by $M / m_{V}$. This is in accordance with the Goldstone boson equivalence theorem, where the vertex involving the longitudinal gauge boson simply restores to that of the Goldstone bosons given by $y\mu H E_1^c$, which governs the VLL-gauge boson coupling.

The mass eigenstates of heavy leptons can be subsequently obtained through another rotation\footnote{The left- and right-handed fields share the same rotation matrix $R$ at leading order in $\epsilon$.}
\begin{equation}
\hspace{-0.5em}
    \begin{bmatrix}
        E_{1,m} \\ E_{2,m} \\ E_{3,m}
    \end{bmatrix}
    =
    R\!\!
    \begin{bmatrix}
        E_{1,f} \\ E_{2,f} \\ E_{3,f}
    \end{bmatrix}\!\!, \hspace{0.2em}\mbox{with } R = \begin{bmatrix}
        \frac{1}{2} & -\frac{1}{\sqrt{2}} & \frac{1}{2} \\ -\frac{1}{\sqrt{2}} & 0 & \frac{1}{\sqrt{2}} \\ \frac{1}{2} & \frac{1}{\sqrt{2}} & \frac{1}{2}
    \end{bmatrix}, \label{eq:mixingV}
\end{equation}
with split masses
\begin{equation}
    M_{E_1} = M(\scalebox{0.9}{$1-$} \scalebox{1.1}{$\frac{\epsilon}{\sqrt{2}}$}),\hspace{0.5em} 
    M_{E_2} = M,\hspace{0.5em} 
    M_{E_3} = M(\scalebox{0.9}{$1+$} \scalebox{1.1}{$\frac{\epsilon}{\sqrt{2}}$}) .
    \label{eq:mass}
\end{equation} 
For the case of electron flavor, the mixed interaction Eq.~(\ref{eq:mixedgauge}) is identical. There are five mass eigenstates, given by
\begin{equation}
    M_{E_{1,5}} = M (\scalebox{0.9}{$1\mp$}\scalebox{1.1}{$\frac{\sqrt{3}}{2}$}\epsilon),\hspace{0.5em} M_{E_{2,4}} = M (\scalebox{0.9}{$1\mp$}\scalebox{1.1}{$\frac{1}{2}$}\epsilon),\hspace{0.5em} M_{E_3} = M. \notag
\end{equation}
Allowing different diagonal and hopping terms in the mass matrix Eq.~(\ref{eq:Mmatrix2}) would generally modify both the mixing angles and the mass splittings.

\subsection{Cross Section and Decay Width}
\label{sec:xsec}

For VLL pair production induced by their universal gauge interactions, the couplings to the three mass eigenstates are equal. 
The cross section from the hypercharge gauge boson exchange is given by \footnote{Here and henceforth, $E_i$ denotes the Dirac fermion and $\bar{E}_i$ denotes its antiparticle. We have omitted the subscript $m$ for mass eigenstates} 
\begin{equation}
    \sigma(f\bar{f}\rightarrow E_i\bar{E}_i) = \frac{g_1^4}{24\pi s } \frac{Y_{f_L}^2+Y_{f_R}^2}{N_c} \beta 
    \Big(1 +\frac{2M_{E_i}^2}{s}\Big) , 
    \label{eq:X}
\end{equation}
where $\beta=\sqrt{1-4M_{E_i}^2/s}$, $Y$ and $N_c=3$ are the hypercharge and color factor. It is important to note that due to the multiplicity of $D$ nearly degenerate states in the chain model, a simultaneous production of $D$ VLLs would be expected, rendering potentially spectacular signals to search for.

For single VLL production induced by mixing with the muon, it directly probes the parameters in the underlying theory. First, the cross section of a single VLL production is governed by the longitudinal gauge boson and thus directly measures the Yukawa coupling of the chain model
\begin{equation}
    \sigma(X\rightarrow E_{1,f}) \propto y^2, 
\end{equation}
as discussed in Sec.~\ref{sec:mass and mixing}. Second, the production rate of the three mass eigenstates satisfies
\begin{align}
    \sigma(X\rightarrow E_{i,m}) \simeq |R_{i1}|^2 \sigma(X\rightarrow E_{1,f}),
    \label{eq:R}
\end{align}
which implies
\begin{align}
    \sigma(X\rightarrow E_1):\sigma(X&\rightarrow E_2):\sigma(X\rightarrow E_3) \simeq 1:2:1 , \label{eq:rate ratio}
\end{align}
for the case of equal $M_i$ and equal $m_{ij}$ in Eq.~(\ref{lag1}), up to the small mass difference, resulting in three separate mass peaks with unequal heights. The sum of these three peaks yields the cross section of the first flavor eigenstate. Therefore, measuring the single production rate ratio of different VLL states directly measures the mixing matrix $R$.
The ratio of the single production rates in Eq.~(\ref{eq:rate ratio}) is the prediction of the mass mixing in the heavy sector in Eq.~(\ref{eq:Mmatrix2}).

Upon production, the heavy VLLs promptly decay to an SM lepton plus gauge boson or Higgs.  Naively, the longitudinal gauge bosons in the final state lead to the energy-enhanced decay rate $\sim M^3/v^2.$ However, taking into account the lepton mixing $\lambda \sim yv/M$, the resultant decay width of the three mass eigenstates is 
\begin{equation}
    \Gamma(E_i)\simeq \frac{1}{16\pi}|R_{i1}|^2y^2M . 
\end{equation}
Same as in Eqs.~(\ref{eq:R}) and (\ref{eq:rate ratio}), one also has
\begin{equation}
    \Gamma(E_1) : \Gamma(E_2) : \Gamma(E_3) \simeq 1:2:1 . 
    \label{eq:Rgamma}
\end{equation} 
As discussed in Sec.~\ref{sec:mass and mixing}, the heavy VLL dominantly decays to longitudinal gauge bosons. 
The Goldstone boson equivalence theorem predicts the branching ratios of $\mathrm{Br}(E_i\rightarrow \mu h):\mathrm{Br}(E_i\rightarrow \mu Z):\mathrm{Br}(E_i\rightarrow \nu W)\simeq 1:1:2$ for $M\gg v$. 

The discussion above equally applies to electron flavor, whose ratio of single production rates for the five VLL states, analogous to Eqs.~(\ref{eq:rate ratio}) and (\ref{eq:Rgamma}), is $1:3:4:3:1$.

Beyond the above simple setting of $M_i=M$ and $m_{ij}=m=\epsilon M$, a different mixing matrix with unequal diagonal or hopping terms (cf.~Eq.~(\ref{eq:Mmatrix2})) would lead to different cross section predictions, making it a sensitive probe of the underlying mass matrix from the mixing among the VLL multiplets (cf.~Eq.~(\ref{eq:mixingV})). In the other limiting case when $m_{ij}\ll |M_i-M_j|$, the mixing would be subdominant and the rotation from flavor to mass eigenstates is trivial at leading order. The VLL chain production would be dominated by one leading peak corresponding to $E_{1,f}$.

\section{VLL Chain at Hadron Colliders}

Adopting the MadGraph package \cite{Alwall:2011uj}, we perform the parton-level calculation and analyses for the VLL signals and the SM backgrounds. We impose suitable kinematic cuts to simulate the detector acceptance, but do not carry out hadronic level detector simulations.
\subsection{Production Rates}

\begin{figure}[tb]
    \centering
    \hspace{1em}
    \begin{minipage}[c]{0.45\columnwidth}
    \begin{tikzpicture}[
        scale=0.6,
        thick,
        decoration={
            markings,
            mark=at position 0.60 with {\arrow{Latex}}
        }
    ]
    \coordinate (L) at (0.3,0);
    \coordinate (R) at (2.1,0);
    \draw[postaction={decorate}] (-0.7,1.4) -- (L);
    \draw[postaction={decorate}] (L) -- (-0.7,-1.4);
    \draw[
        decorate,
        decoration={snake,amplitude=0.7mm,segment length=3mm}
    ] (L) -- (R);
    \draw[postaction={decorate}] (R) -- (3.1,1.4);
    \draw[postaction={decorate}] (3.1,-1.4) -- (R);
    \node at (-1.8,2.2) {(a)};
    \node at (-1.6,-2.2) {$\phantom{a}$};
    \node at (-1.0,1.6) {$q$};
    \node at (-1.1,-1.6) {$q^( {'} \hspace{-0.5em}\phantom{q}^) $};
    \node at (1.2,0.45) {$\gamma/Z/W$};
    \node at (4.0,1.55) {$E_i, \mu, \nu$};
    \node at (3.5,-1.55) {$\bar{E}_i$};
    \end{tikzpicture}
    \end{minipage}
    \hspace{0.5em}
    \begin{minipage}[c]{0.45\columnwidth}
    \begin{tikzpicture}[
        scale=0.6,
        thick,
        decoration={
            markings,
            mark=at position 0.7 with {\arrow{Latex}}
        }
    ]
    \coordinate (A) at (0,1.5);
    \coordinate (B) at (0,0.45);
    \coordinate (C) at (0,-0.45);
    \coordinate (D) at (0,-1.5);
    \draw[postaction={decorate}]
        (-1.7,2.2) -- (A);
    \draw[postaction={decorate}]
        (A) -- (1.7,2.2);
    \draw[postaction={decorate}]
        (D) -- (-1.7,-2.2);
    \draw[postaction={decorate}]
        (1.7,-2.2) -- (D);
    \draw[
        decorate,
        decoration={snake,amplitude=0.7mm,segment length=2mm}
    ] (A) -- (B);

    \draw[
        decorate,
        decoration={snake,amplitude=0.7mm,segment length=2mm}
    ] (C) -- (D);
    \draw[
        postaction={
            decorate,
            decoration={
                markings
            }
        }
    ] (B) -- (C);
    \draw[postaction={decorate}]
        (B) -- (1.65,0.65);
    \draw[postaction={decorate}]
        (1.65,-0.65) -- (C);

    \node at (-2.8,2.2) {(b)};
    \node at (-1.95,2.2) {$q$};
    \node at (2.3,2.35) {$q^( {'} \hspace{-0.5em}\phantom{q}^) $};
    \node at (-2.1,-2.0) {$q^( {''} \hspace{-0.5em}\phantom{q}^) $};
    \node at (2.4,-2.1) {$q^( {'''} \hspace{-0.5em}\phantom{q}^) $};
    \node at (2,0.70) {$E_i$};
    \node at (2.2,-0.65) {$\mu/\nu$};
    \node[left] at (-0.15,0.95) {$V$};
    \node[left] at (-0.15,-0.95) {$V$};
    \end{tikzpicture}
    \end{minipage}

    \caption{Feynman diagrams for (a) the pair and single VLL production from $q\bar q$ annihilation, and (b) single production $E_i\mu, E_i\nu$ via vector boson fusion (VBF) at hadron colliders.}
    \label{fig:feyn diag pp}
\end{figure}
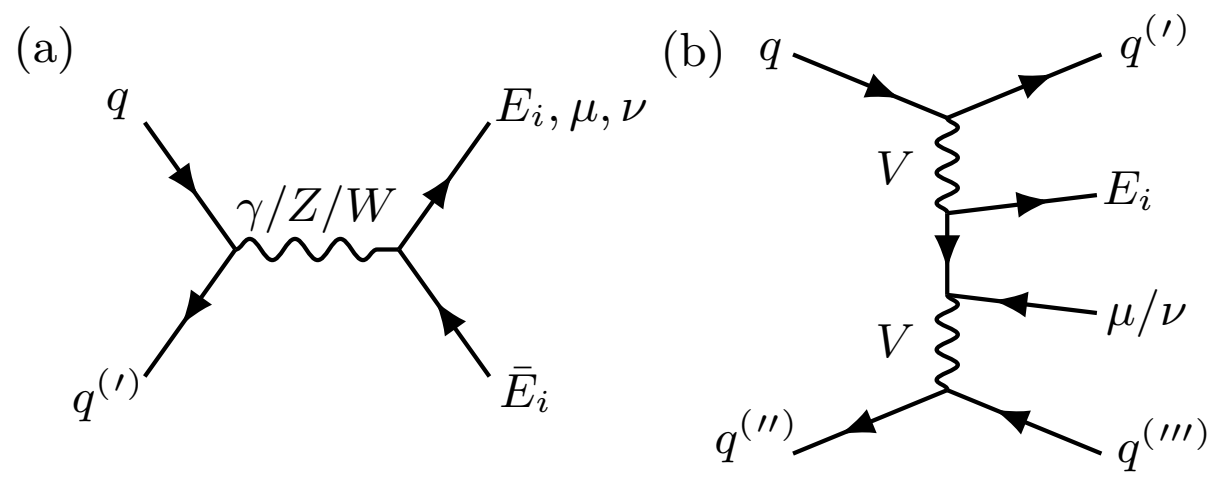

\begin{figure}[tb]
    \centering
    \includegraphics[width=0.9\linewidth]{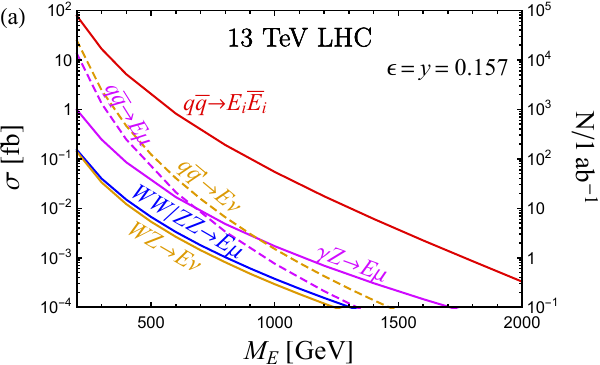} 
    \includegraphics[width=0.9\linewidth]{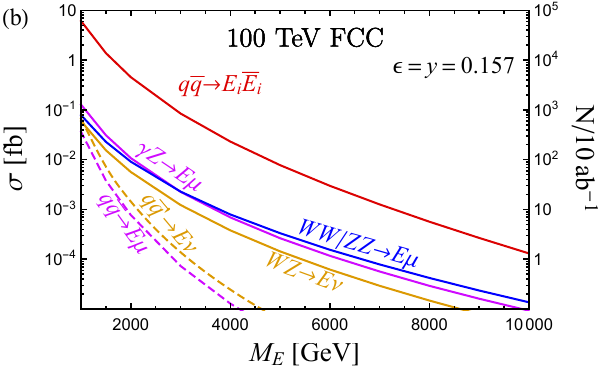} 
    \caption{Cross sections and event rates (right vertical axis) for leading VLL production processes as a function of VLL mass (a) at LHC and (b) FCC-hh.} 
    \label{fig:LHC}
\end{figure}

In hadronic collisions, the leading VLL pair production is through quark annihilation via gauge interaction, as shown in Fig.~\ref{fig:feyn diag pp}(a) and given in Eq.~(\ref{eq:X}).  Single VLL production is induced by mixing through direct annihilation $q \bar q \to Z^*\to E \mu$  and $q \bar q' \to W^*\to E \nu$ as shown in Fig.~\ref{fig:feyn diag pp}(a), and the VBF process as shown in Fig.~\ref{fig:feyn diag pp}(b).
The production cross sections for the leading processes before any acceptance cuts are shown in Fig.~\ref{fig:LHC}(a) for the LHC and Fig.~\ref{fig:LHC}(b) for the FCC-hh. For illustration, we use the benchmark parameters 
\begin{equation}
    y=\epsilon=(y_\mu)^{\frac{1}{4}} = 0.157,
\end{equation}
unless stated otherwise.

At LHC energy, pair production for $E_1\bar E_1$ produces $\sim 150$ events at $M_{E_1}=1\,\mathrm{TeV}$ at $3\,\mathrm{ab}^{-1}$. Because of the presence of multiple VLL states, the total cross section is multiplied by 3 for muon-flavored VLL, and by 5 for electron-flavored VLL, modulo kinematic factors, yielding a sizable production rate. 
The single production is dominated by $q\bar{q} \to Z^*/W^* \rightarrow E\mu/\nu$ at low mass $M_E<700$ GeV and by $\gamma Z\to E\mu$ at high mass. However, the production rate is about one to two orders of magnitude smaller than the pair production, with no multiplicity enhancement.

FCC-hh would greatly increase the production cross section and thus substantially extend the mass reach, as shown in Fig.~\ref{fig:LHC}(b). The pair production yields around 50 $E_1\bar E_1$ events at $M_{E_1}=5\,\mathrm{TeV}$ at a luminosity of $10\,\mathrm{ab}^{-1}$. For single production, due to the enhanced beam energy, the $WW$ and $ZZ$ fusion processes become comparable to that of $\gamma Z$, and produce around 100 $E_1\mu$ events at $M_E=2.5\,\mathrm{TeV}$ at $10\,\mathrm{ab}^{-1}$.

\subsection{Discovery Potential at the LHC and FCC-hh }

The current searches for VLL at the LHC place a mass bound on the singlet VLL as $M_E>400\,\mathrm{GeV}$ for muon flavor and $M_E>320\,\mathrm{GeV}$ for electron flavor \cite{ATLAS:2024mrr}. We carry out optimal cut-based studies and perform detailed phenomenological analyses of the projected sensitivity in HL-LHC and FCC-hh. 

The VLL production can be reconstructed from its decay to $Z\mu$, $h\mu$, and $W\nu$. Since $M_E \gg m_W$, the two-body kinematics yields highly energetic decay products
\begin{equation}
    E_{\mathrm{lepton}} \approx E_{W,Z,h} \approx {M_E\over 2} , 
    \label{eq:E}
\end{equation}
in the VLL rest frame, which consequently leads to the Jacobian peak in the transverse momentum distribution in the lab frame. 
The $W,Z,h$ subsequently decay to leptons and hadrons, generating various decay channels with multiple leptons and jets. For pair production, the representative channels and branching fractions are shown in Table~\ref{tab:BR}, where we include $h\rightarrow b \bar{b},\, h\rightarrow WW^*$ and the leptonic and hadronic decay of $W,Z$.

In the following, we present the results for two leading channels to illustrate the characteristics and the search strategy for the selection of the signal and background, namely $E\bar E\to (Z\mu)(Z\mu)\rightarrow \mu\mu j j j j$ and $(Z\mu)(W\nu)\rightarrow \mu\nu j j j j$. Including the other decay channels in Table \ref{tab:BR} in a combined analysis would improve the sensitivity reach.

\begin{table}[]
    \centering
    \begin{tabular}{l|c}
        Channel & Branching fraction \\
        \hline
        $1\ell+4J+E_{\mathrm{miss}}$ & 21.4\% \\
        $2\ell+2J+E_{\mathrm{miss}}$ & 6.8\% \\
        $2\ell+4J$ & 10.2\% \\
        $3\ell+2J+E_{\mathrm{miss}}$ & 1.1\% \\
        $4\ell+E_{\mathrm{miss}}$ & 0.4\% \\
        $4\ell+2J$ & 1.1\% \\
        $4\ell+2J+E_{\mathrm{miss}}$ & 0.23\% \\
        \hline
        $1\ell+1\tau+2J+E_{\mathrm{miss}}$ & 3.5\% \\
        $1\ell+2\tau+2J+E_{\mathrm{miss}}$ & 1.8\% \\
        $1\ell+1\tau+4J+E_{\mathrm{miss}}$ & 0.5\% \\
        $2\ell+2\tau+E_{\mathrm{miss}}$ & 0.6\% \\
        $2\ell+2\tau+2J$ & 1.5\% \\
        $2\ell+1\tau+2J+E_{\mathrm{miss}}$ & 0.5\% \\
        $2\ell+1\tau+4J+E_{\mathrm{miss}}$ & 0.5\% 
    \end{tabular}
    \caption{Branching fractions for the decay channels of a pair of VLLs, where $\ell=e,\mu$ and $J=j,j_b$ include both light jets and $b$-jets. $E_{\mathrm{miss}}$ represents the presence of one or more neutrinos.}
    \label{tab:BR}
\end{table}

\begin{figure}[htb!]
    \centering
    \includegraphics[width=0.8\linewidth]{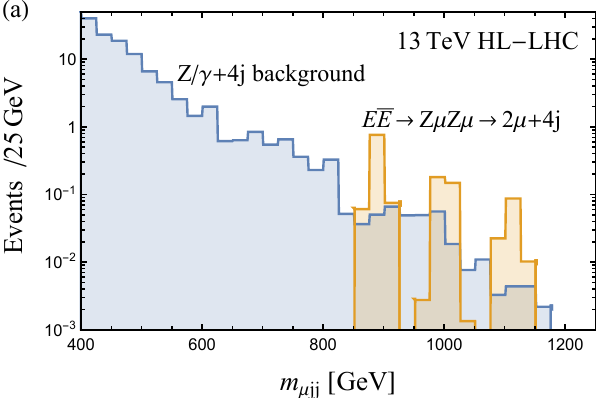} 
    \\[2mm]
    \includegraphics[width=0.82\linewidth]{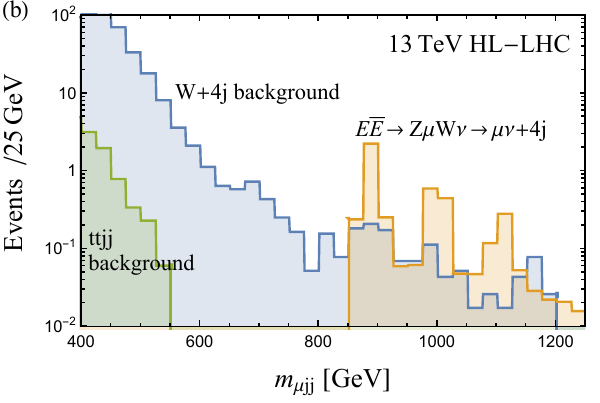} 
    \caption{Multiplet VLL signals and background distributions of the reconstructed invariant mass $m_{\mu j j}$ for $M=1\,\mathrm{TeV}$ at the LHC with 3 ab$^{-1}$ luminosity for (a) $E \bar{E} \rightarrow (Z\mu)(Z\mu)\rightarrow \mu\mu j j j j $, and (b) $E \bar{E} \rightarrow (Z\mu)(W\nu)\rightarrow \mu\nu j j j j $.
    }
    \label{fig:ZmuZmu}
\end{figure}

We first show the $E \bar{E} \rightarrow (Z\mu)(Z\mu)\rightarrow \mu\mu j j j j $ channel, which corresponds to a branching fraction of $1.9\%$. We require the tagging of exactly 2 muons and 4 light jets. The dominant background is $\gamma^*/Z+4j$ where $Z/\gamma$ goes to a pair of muons \cite{ATLAS:2024mrr}. We first adopt  the basic acceptance cuts
\begin{gather}
    p_T(\ell)>10\,\mathrm{GeV},\hspace{1em} p_T(j)>20\,\mathrm{GeV}, \notag \\
    |\eta(\ell)|<2.5,\hspace{1em} |\eta(j)|<5, \\
    \Delta R(\ell,\ell)>0.4,\hspace{1em}\Delta R(\ell,j)>0.4,\hspace{1em}\Delta R(j,j)>0.4 . \notag
\end{gather}
Next, we apply the following cut to suppress the background from $\gamma^*/Z$
\begin{equation}
    \Delta R(\mu^+,\mu^-)>1.
\end{equation}
We require that there exist four jets that satisfy
\begin{equation}
    |m_{j_1j_2}-m_Z|+|m_{j_3j_4}-m_Z|<30\,\mathrm{GeV},
\end{equation}
and the reconstructed masses for $E$ and $\bar{E}$ in close values
\begin{equation}
    |m_{\mu_1j_1j_2}-m_{\mu_2j_3j_4}|<0.1(m_{\mu_1j_1j_2}+m_{\mu_2j_3j_4}).
\end{equation}
To further purify the signal as motivated by Eq.~(\ref{eq:E}), we subsequently require
\begin{gather}
    p_T(\mu_{1})>0.25\,m_{\mu_1j_1j_2},\hspace{1em} p_T(\mu_{2})>0.25\,m_{\mu_2j_3j_4}, \\
    p_T(Z_{1})>0.2\,m_{\mu_1j_1j_2},\hspace{1em} p_T(Z_{2})>0.2\,m_{\mu_2j_3j_4}, \\
    p_T(j) > 0.05\,\mathrm{max}(m_{\mu_1j_1j_2},m_{\mu_2j_3j_4}).
\end{gather}
The resulting distributions for  $M=1\,\mathrm{TeV}$ are shown in Fig.~\ref{fig:ZmuZmu}(a) for HL-LHC. The multiple peaks are prominently visible and reflect the mass splitting in Eq.~(\ref{eq:mass}). The background is efficiently suppressed and the signal efficiency is $40-45\%$ for the mass between $890-1110 \,\mathrm{GeV}$. This yields a significance of $1.4 \sigma$ at 3 ab$^{-1}$ for the benchmark parameter. The $E \bar{E} \rightarrow (h\mu)(h\mu)\rightarrow \mu\mu b \bar b b \bar b $ channel produces a similar signal and a smaller background if effective $b$-tagging is implemented.

The second channel studied is $E \bar{E} \rightarrow (Z\mu)(W\nu)\rightarrow \mu\nu j j j j $, which corresponds to the branching fraction $9.2\%$. We require the tagging of exactly 1 muon and 4 light jets and large missing energy. With the basic selection and isolation cuts, the dominant backgrounds are $W^{(*)}+4j$ with $W^{(*)}\rightarrow\mu\nu$, and $t\bar{t}jj$ with leptonic decay where one lepton is missing and the b-jets are not tagged. In addition to the same acceptance cuts, we apply the following cut
\begin{equation}
    \Delta \phi(p_T(\mu),p_{T,\mathrm{miss}}) > 1. 
\end{equation}
The hadronic decays of $Z,W$ are reconstructed from requiring the dijet mass to be close to $m_Z$ and $m_W$. We demand
\begin{equation}
    |m_{j_1j_2}-m_Z|+|m_{j_3j_4}-m_W|<30\,\mathrm{GeV}.
\end{equation}
We assign the muon to the pair of jets that produces a larger invariant mass $m_{\mu jj}$ as the reconstructed mass of the VLL, and subsequently require 
\begin{gather}
    p_T(\mu) > 0.25\, m_{\mu jj},\hspace{1em} p_{T,\mathrm{miss}} > 0.25\, m_{\mu jj},
    \\
    p_T(Z)>0.2\,m_{\mu jj},\hspace{1em} p_T(W)>0.2\,m_{\mu jj}, \\
    p_T(j) > 0.05\,m_{\mu jj}.
\end{gather}
The resulting distributions for the reconstructed mass are shown in Fig.~\ref{fig:ZmuZmu}(b). The background of leptonic decay $ttjj$ is estimated by requiring one lepton to be forward with $|\eta|>2.5$ and missing, and assuming a $30\%$ rate for each $b$-jet to be untagged.
We find a suppressed background and a signal efficiency of $30-33\%$ for VLL mass between $890-1110\,\mathrm{GeV}$. The significance is $3.1\sigma$ with 3 ab$^{-1}$ for $M=1\,\mathrm{TeV}$.

The multiple peak structure shown in Fig.~\ref{fig:ZmuZmu} is the key feature in a chain model. The multiplicity reflects the length of the chain and subsequently the power suppression on the lepton mass. For the case of electron flavor, five resonant peaks would be expected, corresponding to the five VLL states for the electron mass generation. 

\begin{figure}[htb!]
    \centering
    \includegraphics[width=0.8\linewidth]{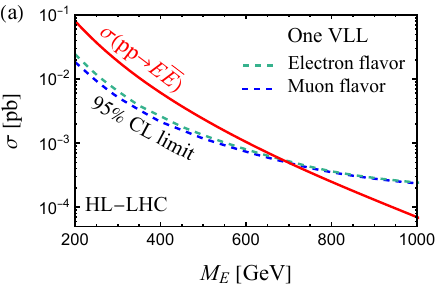} 
    \includegraphics[width=0.8\linewidth]{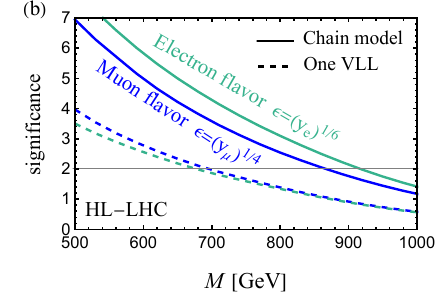} 
    \caption{(a) The 95\% CL exclusion limit on the cross section of one VLL state at HL-LHC, as a function of its mass $M_E$. (b) The statistical significance of VLL observation for 3 ab$^{-1}$ luminosity as a function of the mass parameter $M$. The green and blue solid curves show the significance for electron- and muon-flavored VLL states in the chain model, while the dashed curves show that for a single VLL state. Results are recast from Ref.~\cite{ATLAS:2024mrr,CMS:2024bni}. }
    \label{fig:significance}
\end{figure}

The existing experimental analysis of VLL at LHC covers various decay channels. 
Ref.~\cite{ATLAS:2024mrr} studied $2\ell$, $3\ell$ and $4\ell$ channels, and Ref.~\cite{CMS:2024bni} studied both $3\ell$, $4\ell$ and channels involving $\tau$ leptons. We extrapolate and rescale the $2\ell$ channel sensitivity from~\cite{ATLAS:2024mrr} to HL-LHC and combine it with that of \cite{CMS:2024bni}. The resulting exclusion limit is shown in Fig.~\ref{fig:significance}(a). The combined sensitivity reaches $M_E=700\,\mathrm{GeV}$ at 95\% CL for a single VLL. 

For the chain model in our analyses, due to the multiple VLL states with mass splittings, the signal rate is enhanced. For the case of 3 VLLs for muon flavor and 5 VLLs for electron flavor with equal hopping terms in the Lagrangian, the resulting significance on the mass parameter $M$ is shown in Fig.~\ref{fig:significance}(b), recast from the existing LHC exclusion limit for one VLL. We find a reach of $M=867\,\mathrm{GeV}$ for muon flavor and $M=920\,\mathrm{GeV}$ for electron flavor at $2\sigma$ exclusion at HL-LHC. With the inclusion of the $1\ell$ channel and more advanced machine learning techniques and optimization, the eventual reach in terms of VLL mass and the mass parameter $M$ is expected to be higher than the sensitivity projected here \cite{CMS:2024bni}. 

At FCC-hh, Ref.~\cite{Guedes:2021oqx} shows a mass reach of $M_E=2525\,\mathrm{GeV}$ at 95\% CL for one VLL, using $3\ell$, $4\ell$ and $2\ell+E_{\mathrm{miss}}$ channels. In a chain model, the reach is higher on the mass parameter $M$, as demonstrated above for the HL-LHC.

\section{VLL Chain at High-energy Lepton Colliders} 


High-energy lepton colliders provide a clean experimental environment for opening a new energy threshold and for precision studies of new physics \cite{Roser_2023}. 
In a multi-TeV lepton collider, such as a high-energy $e^+e^-$ linear collider CLIC \cite{CLIC:2018fvx} or a muon collider \cite{InternationalMuonCollider:2025sys}, the leading-order VLL pair production comes from the direct annihilation channel via the hypercharge gauge boson.
The single production induced by mixing is dominated by the effective $Z\mu\rightarrow E_{1,f}$ process, where $Z$ is radiated from lepton, dominantly from the longitudinal gauge boson $Z_L$ at high energies. The cross section is thus given by a convolution with the $Z_L$ parton distribution function (PDF) from lepton \cite{Dawson:1984gx,Chen:2016wkt,Han:2020uid,Garosi:2023bvq}
\begin{equation}
    \sigma(Z\ell\rightarrow E_{i}) = f_{Z_L} (\mbox{$\frac{M_{E_i}^2}{s}$},M_{E_i}^2) \frac{16\pi^2}{s}\frac{\Gamma(E_i\rightarrow Z \ell)}{M_{E_i}} , 
    \label{eq:ZmuE1}
\end{equation}
where $Z_L$ PDF is given by 
\begin{equation}
    f_{Z_L}(x,Q^2)=\frac{g_Z^2}{8\pi^2}\frac{1-x}{x}\big((Q^Z_{\mu_L})^2+(Q^Z_{\mu_R})^2\big)\frac{Q^2}{Q^2+(1-x)m_Z^2},  \notag
\end{equation}
where $x=M_E^2/s$, $Q^2$ is the factorization scale of order $M_E^2$, and
$Q^Z_{\mu_L}=-\frac{1}{2}+s_W^2, Q^Z_{\mu_R}=s_W^2$. There is no logarithmic evolution of the form $\log(Q^2/m_Z^2)$ for longitudinal PDFs at the leading order, due to the ``ultra-collinear'' behavior of spontaneous symmetry breaking \cite{Chen:2016wkt}. The cross section is thus insensitive to the choice of the factorization scale. 
Eq.~(\ref{eq:ZmuE1}) is a good approximation 
for $\sqrt{s} \gg m_Z$ and $x>m_Z^2/Q^2$. Due to the enhancement of the PDF at small $x$, i.e. $\sqrt{s}\gg M_E$, the cross section approaches a constant at high energy
\begin{equation}
    \sigma(Z\ell\rightarrow E_{i}) \sim 2 g_Z^2\big( (Q^Z_{\mu_L})^2+(Q^Z_{\mu_R})^2 \big) \frac{\Gamma(E_i\rightarrow Z \ell)}{M_{E_i}^3} .
\end{equation}
Measurement of the single production rate therefore directly measures the partial decay width of VLL, which gives access to both Yukawa coupling $y$ and the mixing matrix $R_{i1}$.

There are further contributions such as $\gamma \ell\rightarrow E_{1,f} Z$, $\gamma \ell\rightarrow E_{1,f} h$. Although suppressed by the additional QED coupling and two-body phase space, it is enhanced by the large factor $1/x\log (s/m_\ell^2)$ from photon PDF, the cross section therefore features a growth of $\log(s)$ with increasing collision energy. Due to Goldstone boson equivalence, $E_{1,f}Z$ and $E_{1,f}h$ production rates are equal at high energy.

\begin{figure}[htb!]
    \centering
    \includegraphics[width=0.9\linewidth]{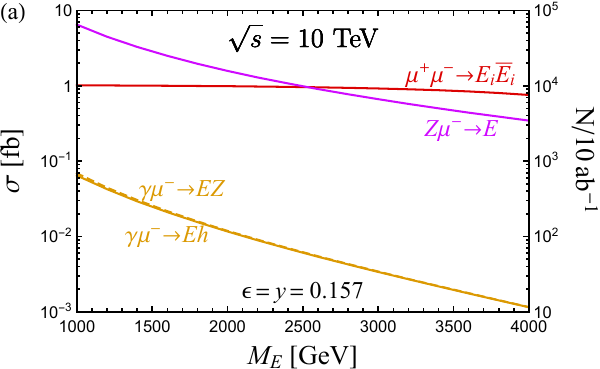} \\[3mm]
    \includegraphics[width=0.9\linewidth]{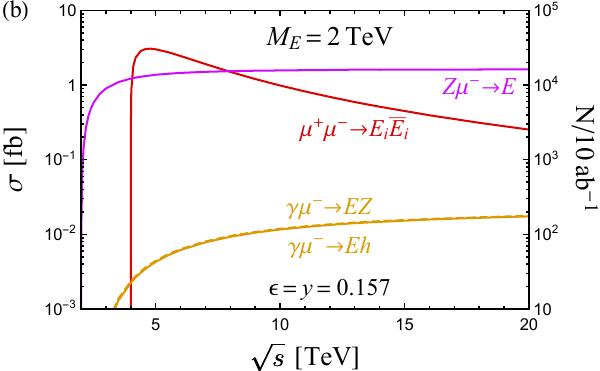} 
    \caption{Cross sections and event rates (right vertical axis) of VLL production at a muon collider, (a) versus the VLL mass for $\sqrt s=10$ TeV, and (b) versus the c.m.~energy for $M_E=2$ TeV.}
    \label{fig:muC}
\end{figure}

Figure \ref{fig:muC} shows the cross sections of the VLL production channels (a) as a function of VLL mass at a 10 TeV muon collider, and (b) as a function of the collision energy for $M_E=2\,\mathrm{TeV}$.  Near and above the threshold $\sqrt s \approx 2M_E$, pair production $\mu^+\mu^-\rightarrow E_i \bar{E}_{i}$ leads to spectacular signal. Single VLL production through $Z\mu\rightarrow E_{1,f}$ extends the reach of the mass in the full kinematic range $\sqrt s \sim M_E$.
These two channels have comparable production rates, while the $\gamma \mu\rightarrow E_{1,f} Z/h$ channel is lower by 1-2 orders of magnitude. 

\begin{figure}[tb]
    \centering
    \includegraphics[width=0.9\linewidth]{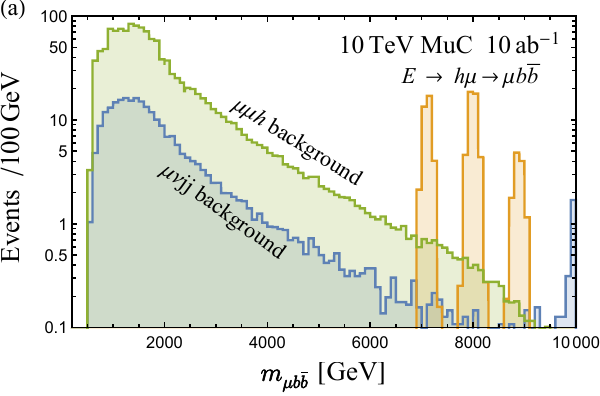} 
    \\[1mm]
    \includegraphics[width=0.885\linewidth]{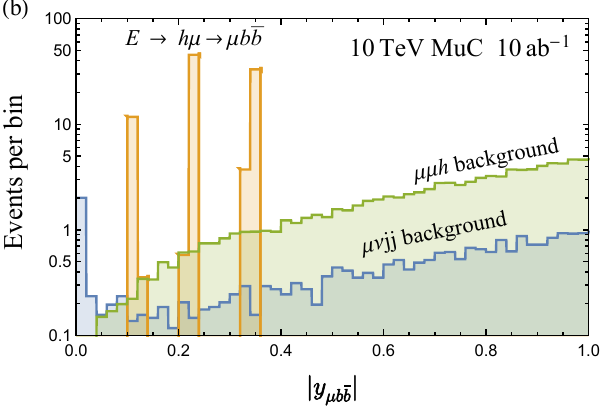} 
    \caption{Multiplet VLL signals and background distributions for $Z\mu \to E$ with $M=8$ TeV at a 10 TeV muon collider (a) for reconstructed mass $m_{\mu b \bar b}$, and (b) rapidity $y_{\mu b \bar b}$.} 
    \label{fig:SB_MuC}
\end{figure}

The leading pair and single production yields a sizable signal with little background.
As a concrete example, we show the signal and background distributions of a single VLL production channel in the chain model. The signal contains a highly forward muon, coming from the beam after radiating a $Z$ boson, with a typical scattering angle $\theta_\mu \approx m_Z/\sqrt s\approx 0.01$, which is usually missing from the detection. The VLL decay produces energetic final states with large transverse momentum $p_T\sim M_E/2$. We consider the $E\rightarrow h\mu\rightarrow \mu b\bar b$ channel for illustration. At high energy, the two $b$-jets are collimated, which could become a single merged jet, denoted by $\mathcal{J}$. We require the tagging of exactly 1 muon and 1 merged jet or 2 jets. The backgrounds are small in a lepton collider, which are primarily obtained from $\mu\mu\rightarrow\mu\nu j j$ and $\mu\mu\rightarrow\mu\mu h$ where one of the muons is missing and $h$ decays to $b\bar{b}$. We apply the following cuts 
\begin{gather}
    p_T(\mu)>10\,\mathrm{GeV},\hspace{1em}|\eta(\mu)|<2.5, \\
    |m_{\mathcal{J}}-m_h|<20\,\mathrm{GeV},\hspace{1em}
    p_{T,\mathrm{miss}} < 100\,\mathrm{GeV},
\end{gather}
and reconstruct the Higgs and VLL, whose mass is given by $m_{\mu b\bar b}$. We further require the following.
\begin{gather}
    p_T(\mu)>0.4 m_{\mu b\bar b},\hspace{1em} p_T(h)>0.4 m_{\mu b\bar b}.
\end{gather}
We note that for a single VLL production without transverse motion, the rapidity takes a fixed value
\begin{equation}
    y_E^{}= {1\over 2}\ln({E_E+p_E^z \over E_E-p_E^z}) = \ln{ \sqrt s\over M_E}.
\end{equation}
Observing the peak in the distribution of $y_E^{}$ would provide another way to determine the resonant mass $M_E$.
The resulting distributions of the VLL multiplets  with $M=8\,\mathrm{TeV}$ and $\epsilon=0.157$  in the signal channel $E\to h\mu \to \mu b\bar b$ in a $10\,\mathrm{TeV}$ muon collider are shown in Fig.~\ref{fig:SB_MuC} for (a) the reconstructed mass $m_{\mu b\bar b}$ and (b) the rapidity $y_{\mu b\bar b}$. 
Due to mixing and mass splitting, the signal features multiple resonant peaks with different production rates. As described in Sec.~\ref{sec:xsec}, measuring their individual rates directly probes the mixing matrix, while the total rate probes the Yukawa coupling $y$.
Therefore, with large production rates and low backgrounds, a lepton collider is ideal to probe both the VLL pair production and the single production that are difficult to observe at hadron colliders, enabling direct experimental measurement of the heavy–light Yukawa couplings and the underlying chain structure.

\begin{figure}[tb]
    \centering
    \includegraphics[width=0.9\linewidth]{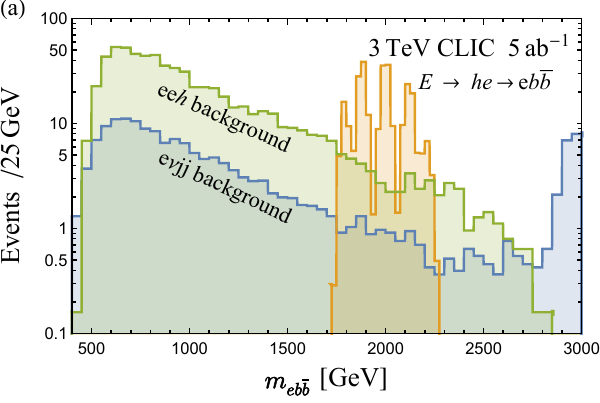} 
    \includegraphics[width=0.885\linewidth]{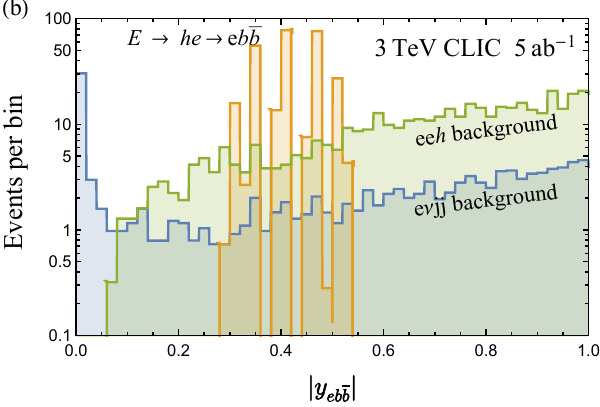} 
    \caption{Multiplet VLL signals and background distributions for $Ze \to E$ with $M=2$ TeV at a 3 TeV CLIC (a) for reconstructed mass $m_{e b \bar b}$, and (b) rapidity $y_{e b \bar b}$.}
    \label{fig:SB_CLIC}
\end{figure}

Similarly, we show the same $m_{eb\bar b}$ and $y_{e b \bar b}$ distributions of electron-flavored VLL multiplets with $M=2\,\mathrm{TeV}$ and $\epsilon=0.12$ at the 3 TeV CLIC in Fig.~\ref{fig:SB_CLIC}, with the same background and cuts on the electrons. The five VLL peaks are quite prominent above the SM backgrounds.

\section{Summary and Conclusions}
\label{sec:con}

In a recent paper \cite{Arkani-Hamed:2026wwy}, a chain flavor model was proposed in order to explain the mass hierarchy and mixing of the fermions in the Standard Model, where chains of vector-like leptons are predicted to have mass near the TeV scale. 
We investigate the production and decay of the VLL multiplets, and quantify the discovery potential at the LHC, FCC-hh, $e^+e^-$ and $\mu^+\mu^-$ colliders.

A characteristic feature of this framework is the presence of multiple VLL states associated with each SM lepton flavor. This gives rise to distinctive resonant signatures from conventional searches for a single VLL. The VLL masses set the new physics scale responsible for fermion mass generation, and the VLL multiplicity determines the length of the chain. Their mass splittings probe the hopping parameters that govern the power-law suppression of the SM lepton masses. Beyond the model-independent pair production channel, single production further measures the underlying interactions in two aspects: the total production rate measures the partial width and hence the heavy–light Yukawa coupling, and the relative rates probe the mixing among the heavy VLL states. Consequently, the observation of multiple resonant peaks would provide a distinctive test of the full underlying chain structure.

Both pair and single production yield sizable rates. The pair production $E\bar E$  is from the model-independent U(1)$_Y$ gauge coupling, which is enhanced due to the presence of multiple VLL states. The single production is dominated by the coupling to longitudinal gauge bosons, which enhances the production rate and gives rise to VBF and vector boson-lepton fusion channels on hadron and lepton colliders. 

In hadron colliders, pair production provides the leading discovery channel, while single VLL production is generally subdominant, as shown in Fig.~\ref{fig:LHC}. 
We demonstrated the multiple reconstructed peaks as the VLL chain in muon flavor at LHC in Fig.~\ref{fig:ZmuZmu}. We find that HL-LHC may probe the chain scale up to $870\,\mathrm{GeV}$ and $920\,\mathrm{GeV}$ for VLLs with muon and electron flavor, respectively. The results are shown in Fig.~\ref{fig:significance}. At FCC-hh, the projected mass reach exceeds $2.5\,\mathrm{TeV}$. 

High-energy lepton colliders offer promising opportunities for both discovering VLL chains and determining their underlying interactions. Both single and pair production provide a large cross section once the collision energy exceeds the threshold, extending the mass reach toward the collider kinematic limit, as shown in Fig.~\ref{fig:muC}. In the benchmark scenario with universal vector-like masses and hopping terms, the predicted single-production rates follow the characteristic ratios $1:2:1$ for the three muon-flavored VLLs and $1:3:4:3:1$ for the five electron-flavored VLLs modulo kinematic factors. We find significant production rates of rich resonant structure at a $\mu^+\mu^-$ collider as seen in Fig.~\ref{fig:SB_MuC}, and at an $e^+e^-$ collider in Fig.~\ref{fig:SB_CLIC}. The multiple signal peaks are spectacularly distinct from the backgrounds.

Our results demonstrate the capability of hadron and lepton colliders in testing the chain flavor mechanism. A combined determination of the multiplicities, mass splittings, and production rates of the VLLs would offer a means of reconstructing the underlying chain structure. Such measurements offer an exciting opportunity and could establish a direct connection between TeV-scale collider phenomena and the origin of the observed fermion mass hierarchy.

\begin{acknowledgments}
We thank Nima Arkani-Hamed, Carolina Figueiredo, Lian-Tao Wang, and Lawrence Hall for stimulating discussions.  
This work was supported in part by the US Department of Energy under grant No.~DE-SC0007914, and in part by Pitt PACC. 
\end{acknowledgments}

\bibliography{ref.bib}

@article{Roser_2023,
   title={On the feasibility of future colliders: report of the Snowmass′21 Implementation Task Force},
   volume={18},
   ISSN={1748-0221},
   url={http://dx.doi.org/10.1088/1748-0221/18/05/P05018},
   DOI={10.1088/1748-0221/18/05/p05018},
   number={05},
   journal={Journal of Instrumentation},
   publisher={IOP Publishing},
   author={Roser, Thomas and Brinkmann, Reinhard and Cousineau, Sarah and Denisov, Dmitri and Gessner, Spencer and Gourlay, Steve and Lebrun, Philippe and Narain, Meenakshi and Oide, Katsunobu and Raubenheimer, Tor and Seeman, John and Shiltsev, Vladimir and Strait, Jim and Turner, Marlene and Wang, Lian-Tao},
   year={2023},
   month=May, pages={P05018} }

@article{Alwall:2011uj,
    author = "Alwall, Johan and Herquet, Michel and Maltoni, Fabio and Mattelaer, Olivier and Stelzer, Tim",
    title = "{MadGraph 5 : Going Beyond}",
    eprint = "1106.0522",
    archivePrefix = "arXiv",
    primaryClass = "hep-ph",
    reportNumber = "FERMILAB-PUB-11-448-T",
    doi = "10.1007/JHEP06(2011)128",
    journal = "JHEP",
    volume = "06",
    pages = "128",
    year = "2011"
}

@article{Chen:2016wkt,
    author = "Chen, Junmou and Han, Tao and Tweedie, Brock",
    title = "{Electroweak Splitting Functions and High Energy Showering}",
    eprint = "1611.00788",
    archivePrefix = "arXiv",
    primaryClass = "hep-ph",
    reportNumber = "PITT-PACC-1602",
    doi = "10.1007/JHEP11(2017)093",
    journal = "JHEP",
    volume = "11",
    pages = "093",
    year = "2017"
}

@article{Buras_2011,
   title={FCNC effects in a minimal theory of fermion masses},
   volume={2011},
   ISSN={1029-8479},
   url={http://dx.doi.org/10.1007/JHEP08(2011)028},
   DOI={10.1007/jhep08(2011)028},
   number={8},
   journal={Journal of High Energy Physics},
   publisher={Springer Science and Business Media LLC},
   author={Buras, Andrzej J. and Grojean, Christophe and Pokorski, Stefan and Ziegler, Robert},
   year={2011},
   month=Aug }

@article{Alonso_2018,
   title={A clockwork solution to the flavor puzzle},
   volume={2018},
   ISSN={1029-8479},
   url={http://dx.doi.org/10.1007/JHEP10(2018)099},
   DOI={10.1007/jhep10(2018)099},
   number={10},
   journal={Journal of High Energy Physics},
   publisher={Springer Science and Business Media LLC},
   author={Alonso, Rodrigo and Carmona, Adrian and Dillon, Barry M. and Kamenik, Jernej F. and Camalich, Jorge Martin and Zupan, Jure},
   year={2018},
   month=Oct }

@article{Arkani-Hamed:1999ylh,
    author = "Arkani-Hamed, Nima and Schmaltz, Martin",
    title = "{Hierarchies without symmetries from extra dimensions}",
    eprint = "hep-ph/9903417",
    archivePrefix = "arXiv",
    reportNumber = "SLAC-PUB-8082",
    doi = "10.1103/PhysRevD.61.033005",
    journal = "Phys. Rev. D",
    volume = "61",
    pages = "033005",
    year = "2000"
}

@article{Giudice:2016yja,
    author = "Giudice, Gian F. and McCullough, Matthew",
    title = "{A Clockwork Theory}",
    eprint = "1610.07962",
    archivePrefix = "arXiv",
    primaryClass = "hep-ph",
    reportNumber = "CERN-TH-2016-223",
    doi = "10.1007/JHEP02(2017)036",
    journal = "JHEP",
    volume = "02",
    pages = "036",
    year = "2017"
}

@article{CLIC:2018fvx,
    author = "de Blas, J. and others",
    collaboration = "CLIC",
    title = "{The CLIC Potential for New Physics}",
    eprint = "1812.02093",
    archivePrefix = "arXiv",
    primaryClass = "hep-ph",
    reportNumber = "CERN-TH-2018-267, CERN-2018-009-M, FERMILAB-TM-2795",
    doi = "10.23731/CYRM-2018-003",
    journal = "CERN Yellow Rep. Monogr.",
    volume = "3",
    pages = "1--282",
    year = "2018"
}

@article{InternationalMuonCollider:2025sys,
    author = "Accettura, Carlotta and others",
    collaboration = "International Muon Collider",
    title = "{The Muon Collider}",
    eprint = "2504.21417",
    archivePrefix = "arXiv",
    primaryClass = "physics.acc-ph",
    reportNumber = "FERMILAB-PUB-25-0309-AD-PPD-T",
    month = "4",
    year = "2025"
}

@article{Froggatt:1978nt,
    author = "Froggatt, C. D. and Nielsen, Holger Bech",
    title = "{Hierarchy of Quark Masses, Cabibbo Angles and CP Violation}",
    reportNumber = "CERN-TH-2519",
    doi = "10.1016/0550-3213(79)90316-X",
    journal = "Nucl. Phys. B",
    volume = "147",
    pages = "277--298",
    year = "1979"
}

@article{Arkani-Hamed:2026wwy,
    author = "Arkani-Hamed, Nima and Figueiredo, Carolina and Hall, Lawrence J. and Manzari, Claudio Andrea",
    title = "{Generating the fermion mass hierarchy at the TeV scale}",
    eprint = "2602.17754",
    archivePrefix = "arXiv",
    primaryClass = "hep-ph",
    month = "2",
    year = "2026"
}

@article{ATLAS:2025cdi,
    author = "Aad, Georges and others",
    collaboration = "ATLAS",
    title = "{Search for Beyond the Standard Model physics with anomaly detection in multilepton final states in pp collisions at $\sqrt{s}=13~\textrm{TeV}$ with the ATLAS detector}",
    eprint = "2508.19778",
    archivePrefix = "arXiv",
    primaryClass = "hep-ex",
    reportNumber = "CERN-EP-2025-189",
    doi = "10.1140/epjc/s10052-025-15218-2",
    journal = "Eur. Phys. J. C",
    volume = "86",
    number = "3",
    pages = "247",
    year = "2026"
}

@article{Cui:2026wwo,
    author = "Cui, Yiheng and Wang, Shiyu and Yu, Zhao-Huan and Zhang, Hong-Hao",
    title = "{Machine learning study on single production of a singlet vectorlike lepton at the Large Hadron Collider}",
    eprint = "2604.11232",
    archivePrefix = "arXiv",
    primaryClass = "hep-ph",
    doi = "10.1103/p8l7-vv3p",
    journal = "Phys. Rev. D",
    volume = "113",
    number = "11",
    pages = "115025",
    year = "2026"
}

@article{ATLAS:2024mrr,
    author = "Aad, Georges and others",
    collaboration = "ATLAS",
    title = "{Search for vector-like leptons coupling to first- and second-generation Standard Model leptons in pp collisions at $ \sqrt{s} $ = 13 TeV with the ATLAS detector}",
    eprint = "2411.07143",
    archivePrefix = "arXiv",
    primaryClass = "hep-ex",
    reportNumber = "CERN-EP-2024-296",
    doi = "10.1007/JHEP05(2025)075",
    journal = "JHEP",
    volume = "05",
    pages = "075",
    year = "2025"
}

@article{ATLAS:2023sbu,
    author = "Aad, Georges and others",
    collaboration = "ATLAS",
    title = "{Search for third-generation vector-like leptons in $pp$ collisions at $\sqrt{s} = 13\,\text{TeV}$ with the ATLAS detector}",
    eprint = "2303.05441",
    archivePrefix = "arXiv",
    primaryClass = "hep-ex",
    reportNumber = "CERN-EP-2022-236",
    doi = "10.1007/JHEP07(2023)118",
    journal = "JHEP",
    volume = "07",
    pages = "118",
    year = "2023"
}

@article{CMS:2022cpe,
    author = "Tumasyan, Armen and others",
    collaboration = "CMS",
    title = "{Search for pair-produced vector-like leptons in final states with third-generation leptons and at least three b quark jets in proton-proton collisions at s=13TeV}",
    eprint = "2208.09700",
    archivePrefix = "arXiv",
    primaryClass = "hep-ex",
    reportNumber = "CMS-B2G-21-004, CERN-EP-2022-153",
    doi = "10.1016/j.physletb.2023.137713",
    journal = "Phys. Lett. B",
    volume = "846",
    pages = "137713",
    year = "2023"
}

@article{ATLAS:2025wgc,
    author = "Aad, Georges and others",
    collaboration = "ATLAS",
    title = "{Search for electroweak production of vector-like leptons in $\tau$-lepton and $b$-jet final states in $pp$ collisions at $\sqrt{s}$ = 13 TeV with the ATLAS detector}",
    eprint = "2503.22581",
    archivePrefix = "arXiv",
    primaryClass = "hep-ex",
    reportNumber = "CERN-EP-2025-055",
    doi = "10.1140/epjc/s10052-025-14748-z",
    journal = "Eur. Phys. J. C",
    volume = "85",
    number = "11",
    pages = "1335",
    year = "2025"
}

@article{CMS:2022nty,
    author = "Tumasyan, Armen and others",
    collaboration = "CMS",
    title = "{Inclusive nonresonant multilepton probes of new phenomena at $\sqrt s$=13{\,}{\,}TeV}",
    eprint = "2202.08676",
    archivePrefix = "arXiv",
    primaryClass = "hep-ex",
    reportNumber = "CMS-EXO-21-002, CERN-EP-2022-008",
    doi = "10.1103/PhysRevD.105.112007",
    journal = "Phys. Rev. D",
    volume = "105",
    number = "11",
    pages = "112007",
    year = "2022"
}

@article{CMS:2024bni,
    author = "Hayrapetyan, Aram and others",
    collaboration = "CMS",
    title = "{Review of searches for vector-like quarks, vector-like leptons, and heavy neutral leptons in proton{\textendash}proton collisions at {\ensuremath{\sqrt{}}}s=13 TeV at the CMS experiment}",
    eprint = "2405.17605",
    archivePrefix = "arXiv",
    primaryClass = "hep-ex",
    reportNumber = "CMS-EXO-23-006, CERN-EP-2024-095",
    doi = "10.1016/j.physrep.2024.09.012",
    journal = "Phys. Rept.",
    volume = "1115",
    pages = "570--677",
    year = "2025"
}

@article{Guedes:2021oqx,
    author = "Guedes, Guilherme and Santiago, Jos{\'e}",
    title = "{New leptons with exotic decays: collider limits and dark matter complementarity}",
    eprint = "2107.03429",
    archivePrefix = "arXiv",
    primaryClass = "hep-ph",
    doi = "10.1007/JHEP01(2022)111",
    journal = "JHEP",
    volume = "01",
    pages = "111",
    year = "2022"
}

@article{Patel:2017pct,
    author = "Patel, Ketan M.",
    title = "{Clockwork mechanism for flavor hierarchies}",
    eprint = "1711.05393",
    archivePrefix = "arXiv",
    primaryClass = "hep-ph",
    doi = "10.1103/PhysRevD.96.115013",
    journal = "Phys. Rev. D",
    volume = "96",
    number = "11",
    pages = "115013",
    year = "2017"
}

@article{AbreudeSouza:2019ixc,
    author = "Abreu de Souza, Fernando and von Gersdorff, Gero",
    title = "{A Random Clockwork of Flavor}",
    eprint = "1911.08476",
    archivePrefix = "arXiv",
    primaryClass = "hep-ph",
    doi = "10.1007/JHEP02(2020)186",
    journal = "JHEP",
    volume = "02",
    pages = "186",
    year = "2020"
}

@article{Ibarra:2017tju,
    author = "Ibarra, Alejandro and Kushwaha, Ashwani and Vempati, Sudhir K.",
    title = "{Clockwork for Neutrino Masses and Lepton Flavor Violation}",
    eprint = "1711.02070",
    archivePrefix = "arXiv",
    primaryClass = "hep-ph",
    reportNumber = "TUM-HEP-1107-17, KIAS-P17115",
    doi = "10.1016/j.physletb.2018.02.047",
    journal = "Phys. Lett. B",
    volume = "780",
    pages = "86--92",
    year = "2018"
}

@article{Mohanta:2026jnq,
    author = "Mohanta, Gurucharan and Patel, Ketan M.",
    title = "{Flavour hierarchies from radiative corrections in latticised theory space}",
    eprint = "2601.10316",
    archivePrefix = "arXiv",
    primaryClass = "hep-ph",
    doi = "10.1007/JHEP07(2026)147",
    journal = "JHEP",
    volume = "07",
    pages = "147",
    year = "2026"
}

@article{Ibarra:2025eya,
    author = "Ibarra, Alejandro and Singh, Aadarsh and Vempati, Sudhir K.",
    title = "{Flavour from fractal mass chains}",
    eprint = "2509.04811",
    archivePrefix = "arXiv",
    primaryClass = "hep-ph",
    doi = "10.1016/j.physletb.2026.140268",
    journal = "Phys. Lett. B",
    volume = "874",
    pages = "140268",
    year = "2026"
}

@article{Altmannshofer:2021qwx,
    author = "Altmannshofer, Wolfgang and Gadam, Sri Aditya",
    title = "{Supersymmetric flavor clockwork model}",
    eprint = "2106.09869",
    archivePrefix = "arXiv",
    primaryClass = "hep-ph",
    doi = "10.1103/PhysRevD.104.035030",
    journal = "Phys. Rev. D",
    volume = "104",
    number = "3",
    pages = "035030",
    year = "2021"
}

@article{Hong:2019bki,
    author = "Hong, Sungwoo and Kurup, Gowri and Perelstein, Maxim",
    title = "{Clockwork Neutrinos}",
    eprint = "1903.06191",
    archivePrefix = "arXiv",
    primaryClass = "hep-ph",
    doi = "10.1007/JHEP10(2019)073",
    journal = "JHEP",
    volume = "10",
    pages = "073",
    year = "2019"
}

@article{vonGersdorff:2020ods,
    author = "von Gersdorff, Gero",
    title = "{Realistic GUT Yukawa couplings from a random clockwork model}",
    eprint = "2005.14207",
    archivePrefix = "arXiv",
    primaryClass = "hep-ph",
    doi = "10.1140/epjc/s10052-020-08764-4",
    journal = "Eur. Phys. J. C",
    volume = "80",
    number = "12",
    pages = "1176",
    year = "2020"
}

@article{Kang:2020cxo,
    author = "Kang, Yoo-Jin and Kim, Soonbin and Lee, Hyun Min",
    title = "{The Clockwork Standard Model}",
    eprint = "2006.03043",
    archivePrefix = "arXiv",
    primaryClass = "hep-ph",
    doi = "10.1007/JHEP09(2020)005",
    journal = "JHEP",
    volume = "09",
    pages = "005",
    year = "2020"
}

@article{Babu:2020tnf,
    author = "Babu, K. S. and Saad, Shaikh",
    title = "{Flavor Hierarchies from Clockwork in SO(10) GUT}",
    eprint = "2007.16085",
    archivePrefix = "arXiv",
    primaryClass = "hep-ph",
    reportNumber = "OSU-HEP-20-10, OSU-HEP-20-09",
    doi = "10.1103/PhysRevD.103.015009",
    journal = "Phys. Rev. D",
    volume = "103",
    number = "1",
    pages = "015009",
    year = "2021"
}

@article{Grossman:1999ra,
    author = "Grossman, Yuval and Neubert, Matthias",
    title = "{Neutrino masses and mixings in nonfactorizable geometry}",
    eprint = "hep-ph/9912408",
    archivePrefix = "arXiv",
    reportNumber = "CLNS-99-1656, SLAC-PUB-8330",
    doi = "10.1016/S0370-2693(00)00054-X",
    journal = "Phys. Lett. B",
    volume = "474",
    pages = "361--371",
    year = "2000"
}

@article{Gherghetta:2000qt,
    author = "Gherghetta, Tony and Pomarol, Alex",
    title = "{Bulk fields and supersymmetry in a slice of AdS}",
    eprint = "hep-ph/0003129",
    archivePrefix = "arXiv",
    reportNumber = "CERN-TH-2000-081, UNIL-IPT-00-06",
    doi = "10.1016/S0550-3213(00)00392-8",
    journal = "Nucl. Phys. B",
    volume = "586",
    pages = "141--162",
    year = "2000"
}

@article{Kaplan:2001ga,
    author = "Kaplan, David Elazzar and Tait, Timothy M. P.",
    title = "{New tools for fermion masses from extra dimensions}",
    eprint = "hep-ph/0110126",
    archivePrefix = "arXiv",
    reportNumber = "SLAC-PUB-9021, ANL-HEP-PR-01-081, EFI-01-44",
    doi = "10.1088/1126-6708/2001/11/051",
    journal = "JHEP",
    volume = "11",
    pages = "051",
    year = "2001"
}

@article{Arkani-Hamed:1999pwe,
    author = "Arkani-Hamed, Nima and Hall, Lawrence J. and Tucker-Smith, David and Weiner, Neal",
    title = "{Flavor at the TeV scale with extra dimensions}",
    eprint = "hep-ph/9909326",
    archivePrefix = "arXiv",
    reportNumber = "LBL-44235, UCB-PTH-99-40, LBNL-44235",
    doi = "10.1103/PhysRevD.61.116003",
    journal = "Phys. Rev. D",
    volume = "61",
    pages = "116003",
    year = "2000"
}

@article{Smolkovic:2019jow,
    author = "Smolkovi{\v{c}}, Aleks and Tammaro, Michele and Zupan, Jure",
    title = "{Anomaly free Froggatt-Nielsen models of flavor}",
    eprint = "1907.10063",
    archivePrefix = "arXiv",
    primaryClass = "hep-ph",
    doi = "10.1007/JHEP10(2019)188",
    journal = "JHEP",
    volume = "10",
    pages = "188",
    year = "2019",
    note = "[Erratum: JHEP 02, 033 (2022)]"
}

@article{Isidori:2010kg,
    author = "Isidori, Gino and Nir, Yosef and Perez, Gilad",
    title = "{Flavor Physics Constraints for Physics Beyond the Standard Model}",
    eprint = "1002.0900",
    archivePrefix = "arXiv",
    primaryClass = "hep-ph",
    doi = "10.1146/annurev.nucl.012809.104534",
    journal = "Ann. Rev. Nucl. Part. Sci.",
    volume = "60",
    pages = "355",
    year = "2010"
}

@article{Silvestrini:2018dos,
    author = "Silvestrini, Luca and Valli, Mauro",
    title = "{Model-independent Bounds on the Standard Model Effective Theory from Flavour Physics}",
    eprint = "1812.10913",
    archivePrefix = "arXiv",
    primaryClass = "hep-ph",
    reportNumber = "CERN-TH-2018-279, UCI-TR-2018-24",
    doi = "10.1016/j.physletb.2019.135062",
    journal = "Phys. Lett. B",
    volume = "799",
    pages = "135062",
    year = "2019"
}

@article{Aebischer:2020dsw,
    author = "Aebischer, Jason and Bobeth, Christoph and Buras, Andrzej J. and Kumar, Jacky",
    title = "{SMEFT ATLAS of $\Delta$F = 2 transitions}",
    eprint = "2009.07276",
    archivePrefix = "arXiv",
    primaryClass = "hep-ph",
    doi = "10.1007/JHEP12(2020)187",
    journal = "JHEP",
    volume = "12",
    pages = "187",
    year = "2020"
}

@article{Calibbi:2012at,
    author = "Calibbi, Lorenzo and Lalak, Zygmunt and Pokorski, Stefan and Ziegler, Robert",
    title = "{Universal Constraints on Low-Energy Flavour Models}",
    eprint = "1204.1275",
    archivePrefix = "arXiv",
    primaryClass = "hep-ph",
    reportNumber = "MPP-2012-70, CERN-PH-TH-2012-085, TUM-HEP-832-12",
    doi = "10.1007/JHEP07(2012)004",
    journal = "JHEP",
    volume = "07",
    pages = "004",
    year = "2012"
}

@article{Ishiwata:2015cga,
    author = "Ishiwata, Koji and Ligeti, Zoltan and Wise, Mark B.",
    title = "{New Vector-Like Fermions and Flavor Physics}",
    eprint = "1506.03484",
    archivePrefix = "arXiv",
    primaryClass = "hep-ph",
    reportNumber = "CALT-TH-2015-029, KANAZAWA-15-07",
    doi = "10.1007/JHEP10(2015)027",
    journal = "JHEP",
    volume = "10",
    pages = "027",
    year = "2015"
}

@article{Leurer:1992wg,
    author = "Leurer, Miriam and Nir, Yosef and Seiberg, Nathan",
    title = "{Mass matrix models}",
    eprint = "hep-ph/9212278",
    archivePrefix = "arXiv",
    reportNumber = "RU-92-59, WIS-92-94-PH",
    doi = "10.1016/0550-3213(93)90112-3",
    journal = "Nucl. Phys. B",
    volume = "398",
    pages = "319--342",
    year = "1993"
}

@article{Greljo:2025mwj,
    author = "Greljo, Admir and Palavri{\'c}, Ajdin and Stefanek, Ben A.",
    title = "{Minimal Flavor Protection for TeV-scale New Physics}",
    eprint = "2512.04159",
    archivePrefix = "arXiv",
    primaryClass = "hep-ph",
    month = "12",
    year = "2025"
}

@article{Morais:2021ead,
    author = "Morais, Ant{\'o}nio P. and Onofre, Ant{\'o}nio and Freitas, Felipe F. and Gon{\c{c}}alves, Jo{\~a}o and Pasechnik, Roman and Santos, Rui",
    title = "{Deep learning searches for vector-like leptons at the LHC and electron/muon colliders}",
    eprint = "2108.03926",
    archivePrefix = "arXiv",
    primaryClass = "hep-ph",
    reportNumber = "CERN-TH-2022-217",
    doi = "10.1140/epjc/s10052-023-11314-3",
    journal = "Eur. Phys. J. C",
    volume = "83",
    number = "3",
    pages = "232",
    year = "2023"
}

@article{Kawamura:2023zuo,
    author = "Kawamura, Junichiro and Shin, Seodong",
    title = "{Current status on pair-produced muon-philic vectorlike leptons in multilepton channels at the LHC}",
    eprint = "2308.07814",
    archivePrefix = "arXiv",
    primaryClass = "hep-ph",
    reportNumber = "CTPU-PTC-23-38",
    doi = "10.1007/JHEP11(2023)025",
    journal = "JHEP",
    volume = "11",
    pages = "025",
    year = "2023"
}

@article{Ashanujjaman:2022cso,
    author = "Ashanujjaman, Saiyad and Choudhury, Debajyoti and Ghosh, Kirtiman",
    title = "{Search for exotic leptons in final states with two or three leptons and fat-jets at 13 TeV LHC}",
    eprint = "2201.09645",
    archivePrefix = "arXiv",
    primaryClass = "hep-ph",
    reportNumber = "IP/BBSR/2021-7",
    doi = "10.1007/JHEP04(2022)150",
    journal = "JHEP",
    volume = "04",
    pages = "150",
    year = "2022"
}

@article{Kumar:2015tna,
    author = "Kumar, Nilanjana and Martin, Stephen P.",
    title = "{Vectorlike Leptons at the Large Hadron Collider}",
    eprint = "1510.03456",
    archivePrefix = "arXiv",
    primaryClass = "hep-ph",
    reportNumber = "FERMILAB-PUB-15-697-T",
    doi = "10.1103/PhysRevD.92.115018",
    journal = "Phys. Rev. D",
    volume = "92",
    number = "11",
    pages = "115018",
    year = "2015"
}

@article{Koren:2021ypn,
    author = {Koren, Seth and {\"O}ktem, Umut},
    title = "{Searching for exotic production of Higgs boson + X to map out new physics}",
    eprint = "2102.06212",
    archivePrefix = "arXiv",
    primaryClass = "hep-ph",
    doi = "10.1103/PhysRevD.104.035033",
    journal = "Phys. Rev. D",
    volume = "104",
    number = "3",
    pages = "035033",
    year = "2021"
}

@article{Cao:2023smj,
    author = "Cao, Qing-Hong and Guo, Jinhui and Liu, Jia and Luo, Yan and Wang, Xiao-Ping",
    title = "{Long-lived searches of vectorlike lepton and its accompanying scalar at colliders}",
    eprint = "2311.12934",
    archivePrefix = "arXiv",
    primaryClass = "hep-ph",
    doi = "10.1103/PhysRevD.110.015029",
    journal = "Phys. Rev. D",
    volume = "110",
    number = "1",
    pages = "015029",
    year = "2024"
}

@article{Baspehlivan:2022qet,
    author = "Baspehlivan, Feyza and Dagli, Burak and Delialioglu, Osman Emre and Sultansoy, Saleh",
    title = "{Why should we search for vector-like leptons?}",
    eprint = "2201.08251",
    archivePrefix = "arXiv",
    primaryClass = "hep-ph",
    month = "1",
    year = "2022"
}

@article{Chala:2020odv,
    author = "Chala, Mikael and Koz{\'o}w, Pawe{\l} and Ramos, Maria and Titov, Arsenii",
    title = "{Effective field theory for vector-like leptons and its collider signals}",
    eprint = "2005.09655",
    archivePrefix = "arXiv",
    primaryClass = "hep-ph",
    doi = "10.1016/j.physletb.2020.135752",
    journal = "Phys. Lett. B",
    volume = "809",
    pages = "135752",
    year = "2020"
}

@article{CMS:2019hsm,
    author = "Sirunyan, Albert M and others",
    collaboration = "CMS",
    title = "{Search for vector-like leptons in multilepton final states in proton-proton collisions at $\sqrt{s}$ = 13 TeV}",
    eprint = "1905.10853",
    archivePrefix = "arXiv",
    primaryClass = "hep-ex",
    reportNumber = "CMS-EXO-18-005, CERN-EP-2019-088",
    doi = "10.1103/PhysRevD.100.052003",
    journal = "Phys. Rev. D",
    volume = "100",
    number = "5",
    pages = "052003",
    year = "2019"
}

@article{L3:2001xsz,
    author = "Achard, P. and others",
    collaboration = "L3",
    title = "{Search for heavy neutral and charged leptons in $e^{+} e^{-}$ annihilation at LEP}",
    eprint = "hep-ex/0107015",
    archivePrefix = "arXiv",
    reportNumber = "CERN-EP-2001-046",
    doi = "10.1016/S0370-2693(01)01005-X",
    journal = "Phys. Lett. B",
    volume = "517",
    pages = "75--85",
    year = "2001"
}

@article{Bordone:2017bld,
    author = "Bordone, Marzia and Cornella, Claudia and Fuentes-Martin, Javier and Isidori, Gino",
    title = "{A three-site gauge model for flavor hierarchies and flavor anomalies}",
    eprint = "1712.01368",
    archivePrefix = "arXiv",
    primaryClass = "hep-ph",
    reportNumber = "ZU-TH-36-17",
    doi = "10.1016/j.physletb.2018.02.011",
    journal = "Phys. Lett. B",
    volume = "779",
    pages = "317--323",
    year = "2018"
}

@article{Fuentes-Martin:2022xnb,
    author = "Fuentes-Martin, Javier and Isidori, Gino and Lizana, Javier M. and Selimovic, Nudzeim and Stefanek, Ben A.",
    title = "{Flavor hierarchies, flavor anomalies, and Higgs mass from a warped extra dimension}",
    eprint = "2203.01952",
    archivePrefix = "arXiv",
    primaryClass = "hep-ph",
    reportNumber = "ZU-TH-08/22",
    doi = "10.1016/j.physletb.2022.137382",
    journal = "Phys. Lett. B",
    volume = "834",
    pages = "137382",
    year = "2022"
}

@article{Garosi:2023bvq,
    author = "Garosi, Francesco and Marzocca, David and Trifinopoulos, Sokratis",
    title = "{LePDF: Standard Model PDFs for high-energy lepton colliders}",
    eprint = "2303.16964",
    archivePrefix = "arXiv",
    primaryClass = "hep-ph",
    doi = "10.1007/JHEP09(2023)107",
    journal = "JHEP",
    volume = "09",
    pages = "107",
    year = "2023"
}

@article{Han:2020uid,
    author = "Han, Tao and Ma, Yang and Xie, Keping",
    title = "{High energy leptonic collisions and electroweak parton distribution functions}",
    eprint = "2007.14300",
    archivePrefix = "arXiv",
    primaryClass = "hep-ph",
    reportNumber = "PITT-PACC 2006",
    doi = "10.1103/PhysRevD.103.L031301",
    journal = "Phys. Rev. D",
    volume = "103",
    number = "3",
    pages = "L031301",
    year = "2021"
}

@article{Dawson:1984gx,
    author = "Dawson, Sally",
    title = "{The Effective W Approximation}",
    reportNumber = "LBL-17497",
    doi = "10.1016/0550-3213(85)90038-0",
    journal = "Nucl. Phys. B",
    volume = "249",
    pages = "42--60",
    year = "1985"
}
\bibliographystyle{apsrev4-1}

\end{document}